\documentclass[a4paper]{article}
\usepackage{graphicx}
\usepackage{amsmath}

\begin{document}

\title{Conversion of entanglement between continuous variable and qubit systems }
\author{Xiao-yu Chen ,Liang-neng Wu , Li-zhen Jiang , Ya-zhuo Li \\
{\small \ Lab. of Quantum Information, China Institute of Metrology,
Hangzhou 310034,China;}}
\date{}
\maketitle

\begin{abstract}
We investigate how entanglement can be transferred between continuous
variable and qubit systems. We find that a two-mode squeezed vacuum state
and a continuous variable Werner state can be converted to the product
states of infinitive number of two-qubit states while keeping the
entanglement. The reverse process is also possible.


\end{abstract}

\section{Introduction}

Quantum information processing (QIP) has been extensively studied for a
qubit system which is a quantum extension of a bit, spanning two-dimensional
Hilbert space. A qubit is realized by a electronic spin, a two-level atom,
the polarization of a photon and a superconductor among others. Parallelly,
much attentions have been paid to the QIP of quantum continuous variable
(CV) system which is a quantum extension of analog information in classical
information theory. CV physical systems such as a harmonic oscillator, a
rotator and a light field are defined in infinitive-dimensional Hilbert
space. While conversions of analog to digital (A/D) and digital to analog
(D/A) are quite usual in information processing, qubit and CV systems are
nearly always treated separately. There have been some pilot works on how to
entangle two separate qubits by an entangled Gaussian field, the efficient
of the transfer is not high \cite{Paternostro}. We would propose a scheme of
perfect transferring the entanglement in this paper.

\section{Entanglement conversion of two mode squeezed vacuum state}

The two-mode squeezed vacuum state $\left| \Psi \right\rangle _{AB}=\sqrt{%
1-\lambda ^2}\sum_{m=0}^\infty \lambda ^m\left| m\right\rangle _A\otimes
\left| m\right\rangle _B$, where $\lambda =\tanh r$ with $r$ the squeezing
parameter. The entanglement of the state is $E(\left| \Psi \right\rangle
)=\cosh ^2r\log \cosh ^2r-$ $\sinh ^2r\log \sinh ^2r$. The interaction
between different systems can cause the transfer of entanglement between the
systems. The scheme of the system considered is that two individual qubits
each interacting with one entangled part of the field. The whole system will
evolve in the way of $\ U\left( t\right) \rho _{AB}\left( 0\right) \otimes
\rho _{CD}\left( 0\right) U^{+}\left( t\right) $ , where $U\left( t\right)
=\exp [-\frac i\hbar (H_{AC}+H_{BD})t]$ is the evolution operator in
interaction picture, and $\rho _{AB}\left( 0\right) =$ $\left| \Psi
\right\rangle _{AB\text{ }AB}\left\langle \Psi \right| $is the initial state
of the CV system while $\rho _{CD}\left( 0\right) =\prod_{k=1}^K\left|
-\right\rangle _{CD\text{ }CD}^{(k)\text{ \quad }(k)}\left\langle -\right| $
is the initial state of the qubit system. Firstly suppose the model
Hamiltonian of entanglement transfer from CV system to qubit system or vice
versa is
\begin{equation}
H_1=\hbar \Omega \left( \sqrt{n}a^{+}\sigma _{-}+a\sqrt{n}\sigma _{+}\right)
,  \label{wave1}
\end{equation}
where $a$ and $a^{+}$ are the photon annihilation and generation operators
respectively, $n=a^{+}a$, $\sigma _{-}$ and $\sigma _{+}$ are operators
which convert the atom (the qubit) from its excited state $\left|
+\right\rangle $ to ground state $\left| -\right\rangle $ and from ground
state to excited state respectively. The Hamiltonian (\ref{wave1}) can be
considered as a kind of nonlinear Jaynes-Cummings model\cite{Mista} \cite
{Chen}. Then $\exp [-\frac i\hbar H_1t_1]\left| m,-\right\rangle =\cos
(m\Omega t_1)\left| m,-\right\rangle -i\sin (m\Omega t_1)\left|
m-1,+\right\rangle .$ If the interaction time $t_1$ is adjusted in such a
way that $\Omega t_1=\pi /2$ then $\exp [-\frac i\hbar H_1t_1]\left|
2m,-\right\rangle =(-1)^m\left| 2m,-\right\rangle $ and $\exp [-\frac i\hbar
H_1t_1]\left| 2m+1,-\right\rangle =-i(-1)^m\left| 2m,+\right\rangle .$ Apply
the evolution operator $U_1(t_1)=\exp [-\frac i\hbar (H_{1AC}+H_{1BD})t_1]$
to the state$\left| \Psi \right\rangle _{AB}\left| --\right\rangle _{CD\text{
}}^{(1)}$ , we have
\begin{equation}
U_1(t_1)\left| \Psi \right\rangle _{AB}\left| --\right\rangle _{CD\text{ }%
}^{(1)}=\left| \Psi \right\rangle _{AB}^{(1)}\left| \Phi \right\rangle
_{CD}^{(1)}
\end{equation}
with $\left| \Psi \right\rangle _{AB}^{(1)}=\sqrt{1-\lambda ^4}%
\sum_{m=0}^\infty \lambda ^{2m}\left| 2m\right\rangle _A\left|
2m\right\rangle _B$ and $\left| \Phi \right\rangle _{CD}^{(1)}=\frac 1{\sqrt{%
1+\lambda ^2}}(\left| --\right\rangle _{CD}^{(1)}-\lambda \left|
++\right\rangle _{CD}^{(1)}).$ It should be noticed that the state after
evolution is a product state of CV system state and two qubit state. The CV
state $\left| \Psi \right\rangle _{AB}^{(1)}$ has even number of photons in
each mode. We can separate the two qubit state $\left| \Phi \right\rangle
_{CD}^{(1)}$ from the combined state, then append another vacuum two qubit
state $\left| --\right\rangle _{CD}^{(2)}$of $CD$ partite to state $\left|
\Psi \right\rangle _{AB}^{(1)}$ , the new state will be $\left| \Psi
\right\rangle _{AB}^{(1)}\left| --\right\rangle _{CD\text{ }}^{(2)}.$ We
would design another interaction Hamiltonian to assign the entanglement of
CV state to two qubit state. The Hamiltonian will be $H_2=\hbar \Omega
\left( \sqrt{n}a^{+}\frac 1{\sqrt{n}}a^{+}\sigma _{-}+a\frac 1{\sqrt{n}}a%
\sqrt{n}\sigma _{+}\right) ,$the evolution will be $U_2(t_2)\left| \Psi
\right\rangle _{AB}^{(1)}\left| --\right\rangle _{CD\text{ }}^{(2)}=\left|
\Psi \right\rangle _{AB}^{(2)}\left| \Phi \right\rangle _{CD}^{(2)}$ with
the interaction time $t_2$ $=\pi /(4\Omega )$, and $\left| \Psi
\right\rangle _{AB}^{(2)}=\sqrt{1-\lambda ^8}\sum_{m=0}^\infty \lambda
^{4m}\left| 4m\right\rangle _A\left| 4m\right\rangle _B$ , $\left| \Phi
\right\rangle _{CD}^{(2)}=\frac 1{\sqrt{1+\lambda ^4}}(\left|
--\right\rangle _{CD}^{(2)}-\lambda ^2\left| ++\right\rangle _{CD}^{(2)}).$
Then we move from the second two qubit to the vacuum state of the third two
qubit of $CD$ partite and so on. The $k-th$ Hamiltonian will be $H_k=\hbar
\Omega [n(\frac 1{\sqrt{n}}a^{+})^{2^{k-1}}\sigma _{-}+(a\frac 1{\sqrt{n}%
})^{2^{k-1}}n\sigma _{+}]$ and interaction time $t_k=\pi /(2^k\Omega ).$ The
whole state reads
\begin{equation}
U_k(t_k)\cdots U_2(t_2)U_1(t_1)\left| \Psi \right\rangle _{AB}(\left|
--\right\rangle _{\text{ }}^{(1)}\left| --\right\rangle ^{(2)}\cdots \left|
--\right\rangle ^{(k)})_{CD}=\left| \Psi \right\rangle _{AB}^{(k)}(\left|
\Phi \right\rangle _{\text{ }}^{(1)}\left| \Phi \right\rangle ^{(2)}\cdots
\left| \Phi \right\rangle ^{(k)})_{CD},
\end{equation}
with $\left| \Psi \right\rangle _{AB}^{(k)}=\sqrt{1-\lambda ^{2^{k+1}}}%
\sum_{m=0}^\infty \lambda ^{2^km}\left| 2^km\right\rangle _A\left|
2^km\right\rangle _B$ , $\left| \Phi \right\rangle _{CD}^{(k)}=\frac 1{\sqrt{%
1+\lambda ^{2^k}}}(\left| --\right\rangle _{CD}^{(k)}-\lambda
^{2^{k-1}}\left| ++\right\rangle _{CD}^{(k)}).$ The entanglement transferred
to qubits system is
\begin{eqnarray}
E(\prod_{j=1}^K\left| \Phi \right\rangle _{CD\text{ }}^{(j)})
&=&\sum_{j=1}^KE(\left| \Phi \right\rangle _{CD\text{ }}^{(j)})=%
\sum_{j=1}^K[\log (1+\lambda ^{2^j})-\frac{\lambda ^{2^j}}{1+\lambda ^{2^j}}%
2^j\log \lambda ] \\
&=&\log \frac{1-\lambda ^{2^{K+1}}}{1-\lambda ^2}-(\frac{\lambda ^2}{%
1-\lambda ^2}-\frac{2^K\lambda ^{2^{K+1}}}{1-\lambda ^{2^{K+1}}})\log
\lambda ^2.  \nonumber
\end{eqnarray}
The entanglement remained at the CV system is $E(\left| \Psi \right\rangle
_{AB}^{(K)})=-\log (1-\lambda ^{2^{K+1}})-\frac{2^{K+1}\lambda ^{2^{K+1}}}{%
1-\lambda ^{2^{K+1}}}\log \lambda .$The total entanglement remains unchanged
for each $K$, $E(\prod_{j=1}^K\left| \Phi \right\rangle _{CD\text{ }%
}^{(j)})+E(\left| \Psi \right\rangle _{AB}^{(K)})=E(\left| \Psi
\right\rangle _{AB}).$When $K\rightarrow \infty ,$ $\lambda
^{2^{K+1}}\rightarrow 0$, thus $E(\left| \Psi \right\rangle
_{AB}^{(K)})\rightarrow 0,$ the entanglement transferred to the qubit system
tends to $E(\left| \Psi \right\rangle _{AB})$. The entanglement is perfectly
transferred. The entanglement transfer is depicted in Fig.1 for different
value of receiving qubit pair number $K$.

\section{Reverse conversion of entanglement}

In the reverse conversion, we have the initial state $\left| \phi
_1\right\rangle _{CD}\left| \phi _2\right\rangle _{CD}$ $\cdots \left| \phi
_k\right\rangle _{CD\text{ }}$, where $\left| \phi _i\right\rangle _{CD\text{
}}=a_{00}^i\left| --\right\rangle ^{(i)}+$ $a_{01}^i\left| -+\right\rangle
^{(i)}+$ $a_{10}^i\left| +-\right\rangle ^{(i)}+a_{11}^i\left|
++\right\rangle ^{(i)}$. The process of entanglement transfer is to transfer
firstly the higher qubit pair ($K-th)$ to the CV bipartite state then the
lower. The result of conversion will be $U_1^{+}(t_1)\left| \phi
_1\right\rangle _{CD}$ $U_2^{+}(t_2)\left| \phi _2\right\rangle _{CD}\cdots $
$U_K^{+}(t_K)\left| \phi _K\right\rangle _{CD\text{ }}$ $\left|
00\right\rangle _{AB}=$ $\left| \psi \right\rangle _{AB}$ $%
\prod_{k=1}^K\left| --\right\rangle _{CD}^{(k)},$ where $\left| \psi
\right\rangle _{AB}=$ $\sum_{n_1,\cdots n_K,m_1,\cdots m_K=0}^1$ $%
\prod_{j=1}^K$ $(-1)^{m_{j+1}+n_{j+1}}i^{m_j+n_j}$ $a_{n_jm_j}^j$ $\left|
n_K\cdots n_1,m_K\cdots m_1\right\rangle ,$with $m_j,$ $n_j=0,1$and $%
m_{K+1}=n_{K+1}=0.$ We denoted $n=\sum_{j=1}^Kn_j2^{j-1}$ for later use. The
Entanglement of the state $\left| \psi \right\rangle _{AB}$ is equal to that
of a state $\left| \psi ^{\prime }\right\rangle
=\sum_{n_1,m_1=0}^1i^{m_1+n_1}a_{n_1m_1}^1$ $\left| n_1,m_1\right\rangle
\prod_{j=2}^K(\sum_{n_j,m_j=0}^1$ $(-i)^{m_j+n_j}a_{n_jm_j}^j\left|
n_j,m_j\right\rangle )$, thus it is equal to the sum of entanglements of
qubit pairs. We have $E(\left| \psi \right\rangle
_{AB})=\sum_{j=1}^KE(\left| \phi _j\right\rangle _{CD}).$

The conversion procedure will convert a general qubit pair product state $%
\rho _{CD}^{(1)}\otimes \rho _{CD}^{(2)}\otimes \cdots \otimes \rho
_{CD}^{(K)}$ into a continuous variable state $\rho _{AB}$ while keeping the
entanglement due to local unitary operations. The process of reverse
entanglement conversion is to convert firstly the highest two qubits $(K-th)$
to the CV system then the lower. The combined state will evolve to
\begin{equation}
U_1^{\dagger }(t_1)\{\{U_2^{\dagger }(t_2)\cdots \{U_K^{\dagger }(t_K)\left|
00\right\rangle \left\langle 00\right| \otimes \rho
_{CD}^{(K)}U_K(t_K)\}\cdots \otimes \rho _{CD}^{(2)}U_2(t_2)\otimes \rho
_{CD}^{(1)}\}U_1(t_1),
\end{equation}
where $\left| 00\right\rangle \left\langle 00\right| $ is the initial state
of the field ($A$ and $B$ bipartite). Since
\begin{eqnarray}
\exp (\frac i\hbar H_Kt_K)\left| 0,-\right\rangle _K &=&\left|
0,-\right\rangle _K \\
\exp (\frac i\hbar H_Kt_K)\left| 0,+\right\rangle _K &=&\cos (2^{K-1}\Omega
t_K)\left| 0,+\right\rangle _K+i\sin (2^{K-1}\Omega t_K)\left|
2^{K-1},-\right\rangle _K.  \nonumber
\end{eqnarray}
The evolution time is so chosen that $\cos (2^{K-1}\Omega t_K)=0,$ we choose
$2^{K-1}\Omega t_K=\pi /2$ as before. Then $\exp (\frac i\hbar H_Kt_K)\left|
0,+\right\rangle _K=$ $i\left| 2^{K-1},-\right\rangle _K$. The first step
evolution will be $U_K^{\dagger }(t_K)\left| 00\right\rangle \left\langle
00\right| \otimes \rho _{CD}^{(K)}U_K(t_K)=$ $\rho _{AB}^{(K)}\otimes \left|
--\right\rangle _{CD,CD}^{(K)\text{ }(K)}\left\langle --\right| ,$ where $%
U_K(t_K)=\exp [-\frac i\hbar (H_{KAC}+H_{KBD})t_K]$ as before. The basis of $%
\rho _{AB}^{(K)}$ are $\left| n_K2^{K-1},m_K2^{K-1}\right\rangle ,$ with $%
n_K,m_K=0$ or $1.$ We see that all the information of qubit state $\rho
_{CD}^{(K-1)}$ is transferred to the field, leave the two qubit state a
definite blank state. Moreover, the combined state is a direct product of
the field and two qubit state, thus the $K-th$ qubit pair can be dropped
after the evolution. The next step is to transfer $\rho _{CD}^{(K-1)}\ $to
the remained field $\rho _{AB}^{(K)}$. Since when $2^{K-2}\Omega t_{K-1}=\pi
/2,$ we have The state after the evolution will be $\rho
_{AB}^{(K-1)}\otimes \left| --\right\rangle _{CD,\text{ \quad }CD}^{(K-1)%
\text{ }(K-1)}\left\langle --\right| \otimes \left| --\right\rangle
_{CD,CD}^{(K)\text{ }(K)}\left\langle --\right| .$ The basis of $\rho
_{AB}^{(K)}$ are $\left| n_K2^{K-1}+n_{K-1}2^{K-2},\text{ }%
m_K2^{K-1}+m_{K-1}2^{K-2}\right\rangle ,$ with $n_{K-1},m_{K-1}=0$ or $1.$
The quantum state of a pair of the two qubits $\rho _{CD}^{(K)}$ $\rho
_{CD}^{(K-1)}$ are transferred to $\rho _{AB}^{(K-1)}.$ When all the two
qubits are transferred to the field, we get at last a bipartite quantum CV
state $\rho =\rho _{AB}^{(1)}$ while leaving all the two qubit series in the
lower energy level state $\prod_{k=1}^K\left| --\right\rangle _{CD}^{(k)}.$
Thus the reverse conversion procedure will convert a general qubit pair
product state $\rho _{CD}^{(1)}\otimes \rho _{CD}^{(2)}\otimes \cdots
\otimes \rho _{CD}^{(K)}$ into a continuous variable bipartite state $\rho $
while keeping the entropy of the whole state. This is due to the fact that
local unitary transformation does not change the entanglement. We here
divide the system into $AC$ and $BD$ subsystems. The reverse conversion is
perfect.

\section{Entanglement conversion of continuous variable Werner state}

CV Werner state is a mixture of the two mode squeezed vacuum state and the
two mode thermal state \cite{Mista}
\[
\rho _W=p\rho _{TMSV}+(1-p)\rho _T,\text{ \qquad }0\leq p\leq 1.
\]
Where
\begin{eqnarray*}
\rho _{TMSV} &=&(1-\lambda ^2)\sum_{m,n=0}^\infty \lambda ^{m+n}\left|
mm\right\rangle \left\langle nn\right| , \\
\rho _T &=&(1-v)^2\sum_{m,n=0}^\infty v^{m+n}\left| mn\right\rangle
\left\langle mn\right| .
\end{eqnarray*}
The sufficient conditions of inseparability and separability as well as
other physical properties have been displayed. Denote partial transposed
state of $\rho _W$ as $\rho _W^{T_A}.$ The eigenvalues of $\rho _W^{T_A}$
are \cite{Mista}
\begin{eqnarray*}
x^{(l)} &=&p(1-\lambda ^2)\lambda ^{2l}+(1-p)(1-v)^2v^{2l},\text{ \qquad }%
l=0,1,\cdots \\
x_{1,2}^{(mn)} &=&(1-p)(1-v)^2v^{m+n}\pm p(1-\lambda ^2)\lambda ^{m+n},\text{
\qquad }m\neq n;\text{ }m,n=0,1,\cdots .
\end{eqnarray*}
The logarithmic negativity (LN) as an entanglement measure of the state $%
\rho _W$ can be calculated \cite{Vidal}. The entanglement measured by LN is $%
E_{LN}=\log _2\left\| \mathcal{\rho }_W^{T_A}\right\| _1,$ where $\left\|
A\right\| _1=$Tr$\sqrt{A^{\dagger }A}.$ In the simplest case of $v=\lambda $
(although LN can be worked out for any values of the parameters), the
inseparable condition is $p>(1-\lambda )/2,$we have

\[
E_{LN}=\log _2[p\frac{1+\lambda }{1-\lambda }+(1-p)\frac{1-\lambda }{%
1+\lambda }].
\]

The conversion of the CV Werner state to a serial of two qubits system leads
to the result state
\[
\rho _{WCD}=p\bigotimes_{k=1}^K(\left| \Phi \right\rangle
_{CD,CD}^{(k)}\left\langle \Phi \right|
^{(k)})+(1-p)\bigotimes_{k=1}^K(\varrho _C^{(k)}\otimes \varrho _D^{(k)}).
\]
Where $\varrho _C^{(k)}=\frac 1{1+v^{2^{k-1}}}(\left| -\right\rangle
_{C,C}^{(k)\text{ }(k)}\left\langle -\right| +v^{2^{k-1}}\left|
+\right\rangle _{C,C}^{(k)\text{ }(k)}\left\langle +\right| )$ and $\varrho
_D^{(k)}$ is the similar. For convenience, we denote $\left| -\right\rangle $
and $\left| +\right\rangle $ as $\left| 0\right\rangle $ and $\left|
1\right\rangle ,$ then the basis of $C$ system (as well as $D$ system) can
be denoted as $\left| n_1\cdots n_K\right\rangle _C$ with $n_k=0,1.$ The
basis can be further simplified as $\left| n\right\rangle _C$ with $%
n=\sum_{k=1}^Kn_k2^{k-1}.$ Thus
\[
\rho _{WCD}=p\frac{1-\lambda ^2}{1-\lambda ^{2^{K+1}}}%
\sum_{n,m=0}^{2^K-1}(-1)^{\sum_{k=1}^K(n_k+m_k)}\lambda ^{m+n}\left|
mm\right\rangle \left\langle nn\right| +(1-p)\frac{(1-v)^2}{(1-v^{2^K})^2}%
\sum_{n,m=0}^{2^K-1}v^{m+n}\left| mn\right\rangle \left\langle mn\right| .
\]
The partially transposed matrix $\rho _{WCD}^{T_C}$ has a block diagonal
form with $1\times 1$blocks in one-dimensional subspaces spanned by vectors $%
\{\left| mm\right\rangle \}$, $m=0,1,...$ and $2\times 2$ blocks in
two-dimensional subspaces spanned by vectors $\{\left| mn\right\rangle
,\left| nm\right\rangle ,m\neq n\}$, $m,n=0,1,....$ Consequently, the
eigenvalues of the partially transposed matrix $\rho _{WCD}^{T_C}$ can
easily be calculated as roots of quadratic equations and read
\begin{eqnarray*}
x^{(l)} &=&p\frac{1-\lambda ^2}{1-\lambda ^{2^{K+1}}}\lambda ^{2l}+(1-p)%
\frac{(1-v)^2}{(1-v^{2^K})^2}v^{2l},\text{ \qquad }l=0,1,\cdots \\
x_{1,2}^{(mn)} &=&(1-p)\frac{(1-v)^2}{(1-v^{2^K})^2}v^{m+n}\pm p\frac{%
1-\lambda ^2}{1-\lambda ^{2^{K+1}}}\lambda ^{m+n},\text{ \qquad }m\neq n;%
\text{ }m,n=0,1,\cdots .
\end{eqnarray*}
After the conversion, the entanglement measured by LN for the case of $%
v=\lambda $ will be
\[
E_{LNCD}=\log _2[p\frac{1+\lambda }{1-\lambda }\frac{1-\lambda ^{2^K}}{%
1+\lambda ^{2^K}}+(1-p)\frac{1-\lambda }{1+\lambda }\frac{1+\lambda ^{2^K}}{%
1-\lambda ^{2^K}}].
\]
The entanglement transfer is shown in Fig.2 for different value of receiving
qubit pair number $K$.

\section{Conclusion}

The entanglement of continuous variable system can be converted to the
entanglement of a serial of qubit pairs. The conversion error can be made
arbitary low if the series is long enough. The reverse conversion is perfect
as far as the different qubit pairs are not correlated before the reverse
conversion. For a typical two mode squeezed vacuum state prepared in
laboratary (the squeezing parameter $r<3$), the qubit system with eight
pairs of qubits is sufficiently good in simulation of the original CV\
system. This is indicated by our calculations of the entanglement transfer
of two mode squeezed vaccum state or CV Werner state. In quantum information
theory, the entanglement of two qubits changes from $0$ to $1,$ this
flexibility makes the error of conversion quite small as we can see from the
figures.

Funding by the National Natural Science Foundation of China (Grant No.
10575092,10347119), Zhejiang Province Natural Science Foundation (Grant No.
RC104265) and AQSIQ of China (Grant No. 2004QK38) are gratefully
acknowledged.

\begin{figure}[tbp]
\includegraphics[ trim=0.000000in 0.000000in -0.138042in 0.000000in,
height=2.0081in, width=2.5097in ]{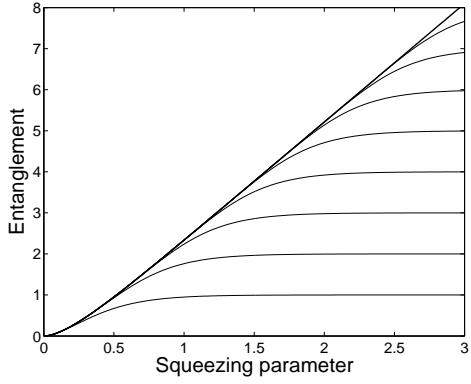} \caption{The thick
line is for CV state, The thin lines from bottom to top are for
the entanglement transferred of K=1,2,...,8 respectively.}
\end{figure}

\begin{figure}[tbp]
\includegraphics[ trim=0.000000in 0.000000in -0.138042in 0.000000in,
height=2.0081in, width=2.5097in ]{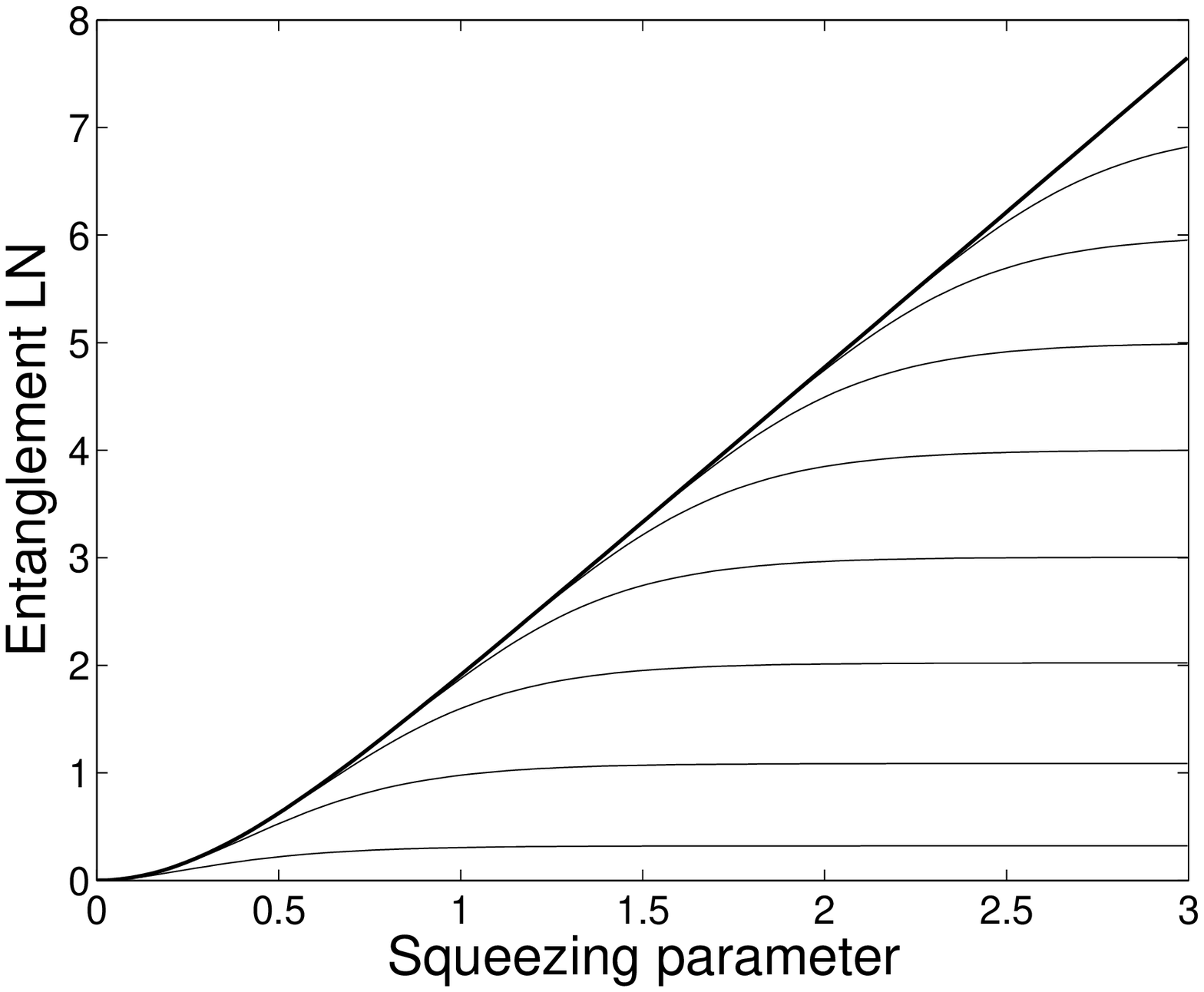} \caption{The thick
line is for CV state, The thin lines from bottom to top are for
the entanglement transferred of K=1,2,...,8 respectively. p=0.5.}

\end{figure}

\end{document}